\documentclass[epj]{svjour}
\usepackage{amsmath}
\usepackage{amssymb}
\usepackage{epsfig}

\newcommand{\be}{\begin{eqnarray}}
\newcommand{\ee}{\end{eqnarray}}

\begin{document}
\title{\hfill {\footnotesize FZJ--IKP(TH)--2007--03} \\
$\Lambda N$ scattering length from the reaction $\gamma d \to K^+\Lambda n$}

\author{A. Gasparyan\inst{1,2}, J. Haidenbauer\inst{3}, 
C. Hanhart\inst{3}, and K. Miyagawa\inst{4}}

\institute{
Institute of Theoretical and Experimental Physics,
117259, B. Cheremushkinskaya 25, Moscow, Russia \and 
Gesellschaft f\"ur Schwerionenforschung (GSI), Planck Str.~1, 
D-64291 Darmstadt, Germany \and 
Institut f\"ur Kernphysik (Theorie), Forschungszentrum J\"ulich,
D-52425 J\"ulich, Germany \and 
Simulation Science Center, Okayama University of Science, 
1-1 Ridai-cho, Okayama 700-0005, Japan
}
\date{Received: date / Revised version: date}

\abstract{
The perspects of utilizing the strangeness-production reaction $\gamma d \to K^+\Lambda n$
for the determination of the $\Lambda n$ low-energy scattering parameters
are investigated. The spin observables that need to be measured in order to 
isolate the $\Lambda n$ singlet ($^1S_0$) and triplet ($^3S_1$) states are identified. 
Possible kinematical regions where the extraction of the $\Lambda n$ scattering lengths
might be feasible are discussed. 
}

\authorrunning{A. Gasparyan et al.} 
\titlerunning{$\Lambda N$ scattering length.}

\maketitle

\section{Introduction}
The experimental information on the $\Lambda N$ interaction at low energies
is rather poor and, moreover, of rather limited accuracy \cite{Alexander}.
Specifically,
the available data do not allow a reliable determination of the $\Lambda N$
low-energy ($^1S_0$, $^3S_1$) scattering parameters. 
Therefore, it has been suggested in the past to consider inelastic
processes where the $\Lambda N$ system is produced in the final
state and to exploit the occuring final-state interaction for the
extraction of those scattering parameters. With this aim in mind, 
we \cite{Gasparyan2003,Gasparyan2005}, but also other 
groups \cite{Bale,Hinter}, have recently looked at the reaction
$pp\to K^+\Lambda p$ which can be studied experimentally at the
COSY facility in J\"ulich and where concrete experiments
have been already performed \cite{C11_1,Bilger,C11_2,Kowina,TOF}. 

In the present paper we want to investigate the perspects of utilizing
another strangeness-production reaction, namely $\gamma d \to K^+\Lambda n$
\cite{Renard,Kerbikov,Adel,Li,Yama,Maxwell,Salam,Miya1,Miya2,Laget}, 
for the determination of the $\Lambda N$ low-energy scattering parameters.
In this case pertinent experiments have been announced already long time ago
at CEBAF \cite{Berman}, but are also possible at ELSA in Bonn \cite{ELSA}, 
at the present JLAB facility \cite{JLAB}, and the future MAMI-C project in 
Mainz \cite{MAMI}. 
In this paper we want to discuss the differences and, in particular, 
the merits but also possible disadvantages of considering the photon-induced 
strangeness production. We also investigate the spin dependence of
the production amplitude and identify those spin-dependent observables
that need to be measured to enable a separation of the $^1S_0$ and $^3S_1$
partial waves. Finally, we present quantitative results within a model calculation
for one of the observables in question in order to demonstrate the kind of
signal one could expect in a concrete experiment. 

In our previous papers \cite{Gasparyan2003,Gasparyan2005} we developed
a method for a quantitative study of the final-state interactions in
production reactions with large momentum transfer such as, e.g.,
$pp\to K^+\Lambda p$ or $pp\to K^0\Sigma^+ p$. In general, the method
can be applied when there is a strong interaction in one of the
produced two-body subsystems, and in addition there are no other
channels with near-by thresholds that couple strongly to that
system. Also, the interaction in the other final two-body subsystems
should be weak.  Then it is possible to reconstruct the elastic
two-body ($\Lambda N$, say) amplitude (or at least its threshold value
-- the scattering length) via the invariant mass dependence of the
production amplitude in the region where the relative $\Lambda N$
momentum is small.  The idea is to separate the different momentum
scales appearing in the problem. In fact there are three scales one
has to deal with: by assumption---we look only at very strong final
state interactions leading to large scattering lengths---a very small
scale given by the inverse of the scattering length, $1/a$, of the
relevant final-state interaction, the inverse range of forces in the
case of the elastic scattering, which is usually larger then the
former scale, and -- the largest scale -- the inverse range of the
production operator. From the point of view of analytical properties
of the amplitude the latter two scales are roughly given by the
corresponding closest left hand singularities.  It is clear then that
in the case of elastic scattering the location of those singularities
is determined by the mass of the exchanged meson, whereas for the
production reaction it is fixed in most cases by the value of the
required momentum transfer $q$~\cite{report}. The production amplitude
itself is free of the left hand singularities of the elastic
amplitude, but has the same right hand cut.
Dispersion theory enables to factorize the left hand singularities from the elastic 
amplitude and to represent the production amplitude as a product 
(see Refs. \cite{Muskhelishvili1953,Omnes1958,Frazer1959})
\begin{eqnarray}
\nonumber
A(s,t,m^2)&=&\exp\left[{\frac1\pi\int_{m_0^2}^\infty\frac{\delta(m' \, ^2)}
{m' \, ^2-m^2-i0}dm' \, ^2}\right] \\
&&\times \Phi(s,t,m^2),
\label{dis0}
\end{eqnarray}
where the exponent contains the full information on the right hand singularities of the 
elastic amplitude and the remaining factor $\Phi$ possesses only left hand 
singularities and, therefore, in case of large momentum-transfer reactions, is only weakly 
dependent on $m^2$, the invariant mass 
of the considered two-body subsystem, e.g. of $\Lambda N$. 
In Eq.~(\ref{dis0}) $\delta$ is the elastic $\Lambda N$ phase shift and
$m_0=m_\Lambda+m_N$.
$s$ and $t$ are the total c.m. energy squared and the 4-momentum transfer
(from one of the initial particles to the kaon) squared, respectively. 
In a more realistic situation when inelastic channels are present at higher energy
(as is the case for $\Lambda N$ due to the opening of the $\Sigma N$ channel) 
one can write down a similar expression where the integration involves only 
the range where the
final state interaction is strong \cite{Gasparyan2003}
\begin{eqnarray}
\nonumber
A(m^2)&=&\exp\left[{\frac1\pi\int_{m_0^2}^{m_{max}^2}\frac{\delta(m' \, ^2)}{m' \, ^2-m^2-i0}dm' \, ^2}\right]
\\
&&\times \tilde \Phi(m^2,m_{max}^2) \label{C_int} \ .
\label{A_def}
\end{eqnarray}
Here $\tilde \Phi(m^2,m_{max}^2)$ is also a slowly varying function of
$m^2$ provided that $\delta$ is sufficiently small in the vicinity of
$m_{max}$.  Neglecting the mass dependence of $\tilde
\Phi(m^2,m_{max}^2)$ the scattering length $a_S$ in a specific partial
wave $S$ can be then expressed in terms of the differential partial
production cross section $\sigma_S$:
\begin{eqnarray}
\nonumber
a_S&=&\lim_{{m}^2\to m_0^2}\frac1{2\pi}\left(\frac{m_\Lambda+m_N}
{\sqrt{m_\Lambda m_N}}\right) \\
\nonumber
&&\times {\bf P}
\int_{m_0^2}^{m_{max}^2}dm' \, ^2
\sqrt{\frac{m_{max}^2-{m}^2}{m_{max}^2-m' \, ^2}}\\
\nonumber
&&\times 
\frac1{\sqrt{m' \, ^2-m_0^2} \ (m' \, ^2-{m}^2)}
\log{\left\{\frac{1}{p'}\left(\frac{d^2\sigma_S}{dm' \, ^2dt}\right)\right\}}
\ . \\
&&
\label{final}
\end{eqnarray}
A detailed analysis of the uncertainties of Eq. (\ref{final}) has
shown that the theoretical error of the extracted value for the
scattering length is of the order of $0.3$ fm
\cite{Gasparyan2003}. Note, however, that a possible influence of
meson-baryon interactions in the other two-body subsystems has not
been explicitly included into this estimate so far---we will do this
below.

In this paper we present results for another strangeness-production process that is 
a possible candidate for the extraction of the $\Lambda N$ scattering length, 
namely $\gamma d \to K^+\Lambda n$.
This reaction satisfies formally the main condition
needed for the dispersion integral method to be applied:
The momentum transfer in this reaction is large compared to 
the typical range of the final state $\Lambda N$ interaction. The required c.m.
momentum of the initial photon to produce the $K\Lambda N$ system at threshold
is around $600$ MeV/c. 
However, in contrast to the $NN$ induced reaction, here a new small scale might
enter the reaction depending on the kinematics: for forward going kaons at 
sufficiently high energy the intermediate nucleon is off--shell
only by the small binding energy of the deuteron before the photon couples. 
Then quasifree production dominates the reaction and the dispersion integral
method cannot be applied anymore. Therefore, one has to impose additional 
kinematical conditions to ensure that quasifree production is not allowed or at 
least strongly suppressed. 

In Sec.~\ref{observables} we consider the spin structure of the reaction amplitude
for $\gamma d \to K^+\Lambda n$ and we derive those spin observables
that need to be measured in order to separate the $\Lambda N$ spin-singlet 
and spin-triplet states. 
In Sec.~\ref{influence} we estimate uncertainties of the extracted $\Lambda N$ 
scattering length that could arise from the interaction in the other
final states ($K\Lambda$, $KN$). 
Concrete results for the spin observable that projects on the spin-triplet
state are presented in Sec.~\ref{results}, based on a model calculation by 
Yamamura et al.~\cite{Yama}. Furthermore, as a test 
we apply the dispersion integral method described above to those specific 
model predictions for extracting the $\Lambda n$ $^3S_1$ scattering length.  
We also discuss issues concerning the kinematical regions 
where experiments should be preferably performed in order to ensure
a reliable determination of the scattering lengths. Specifically, we identify 
the kinematical conditions, where the quasifree production is not allowed or 
strongly suppressed and where then the dispersion integral method can be 
reliably applied.
The paper closes with a brief Summary. 

\section{Spin observables}
\label{observables}
An important issue for the extraction of the low-energy
scattering parameters is the separation of the different spin components
in the $\Lambda N$ system. In Ref. \cite{Gasparyan2003} we have shown
that by measuring specific spin observables in the reaction $NN\to NK\Lambda$
one can project on the production of spin-singlet or spin-triplet states. 
Let us now discuss what observables can be used to disentangle the different 
spin states for the reaction $\gamma d \to K^+\Lambda n$. 

We start from the general form for the matrix element of the process
$\gamma d \to K^+\Lambda n$ :
\begin{eqnarray}
\nonumber
M&=&A(\vec \epsilon_d \cdot \vec \epsilon_\gamma)
+\vec B \cdot (\vec \epsilon_d \times \vec \epsilon_\gamma) \\
&&+C^{(ij)}\left(\vec \epsilon_d\, ^i \vec \epsilon_\gamma\, ^j + \vec \epsilon_d\, ^j
\vec \epsilon_\gamma\, ^i
-\frac23\delta^{ij}(\vec \epsilon_d \cdot \vec \epsilon_\gamma)\right),
\end{eqnarray}
where $\vec \epsilon_\gamma$ and $\vec \epsilon_d$ are the polarization 
vectors of the photon and deuteron, respectively.
If we assume the $\Lambda N$ system to be in an $S$--wave, then we
have only the (normalized) initial momentum $\hat p$ and the outgoing kaon momentum 
$\vec q\, '$ available to construct the structures for the coefficients. 
If the final $\Lambda N$ system is in a spin triplet state we have in
addition $\vec S\, '$, the spin vector of the final state, that has to appear linearly 
in the coefficients $A$, $B$ and $C$. Parity conservation demands that both 
$\hat p$ and $\vec q\, '$ 
appear either in an odd number or in an even number. Thus, we have for the
spin singlet case:
\be
\nonumber
A^s&=&a^s \, , \\
\nonumber
\vec B^s&=&b^s(\vec q\, '\times \hat p) \, , \\
C^{(ij)\, s} &=& c^s_1
  {\hat p}^i{\hat p}^j+c_2q'{}^iq'{}^j+c_3q'{}^i\hat p^j \ .
%+c_3
%(\vec q\, '\times \hat p)^i(\vec q\, '\times \hat p)^j
\label{singlet}
\ee
On the other hand, for the spin triplet final state we get
\be
\nonumber
A^t&=&a^t \vec S\, '\cdot(\vec q\, '\times \hat p) \, , \\
\nonumber
\vec B^t&=&b^t_1\vec S\, ' \, + (\vec S\, '\cdot \hat p)\left(b^t_2 \hat p
+b^t_3 \vec  q\, '\right)
 + b^t_4 \vec  q\, '(\vec S\, '\cdot \vec q \, ') \\
\nonumber
C^{(ij)\, t} &=& \vec S\, '\cdot(\vec q\, '\times \hat p)\left(c^t_1
  {\hat p}^i{\hat p}^j+c^t_2q'{}^iq'{}^j+c^t_3q'{}^i\hat p^j
\right) \\
\nonumber 
&&+ c^t_{4} \vec S\, '{}^i (\vec q\, '\times \hat p)^j
+ (\vec S\, '\times \vec q\, ')^i\left(c^t_5\vec q\, '{}^j+c^t_6\hat p^j\right) \\
&&+ (\vec S\, '\times \hat p)^i\left(c^t_{7}\vec q\, '{}^j+c^t_{8}\hat p^j\right). 
\label{triplet}
\ee
Note that the coefficients $a^s$, $b^s$, etc. are functions of
$(\vec{q}\,')^2$ and $\vec{q}\,'\cdot\vec{\hat{p}}$. 
A significant simplification allowing one to separate different spin states
can be achieved if we assume $\vec q\, '$ to be along the beam direction
(in particular then $\vec B^s$ and $A^t$ vanish). This means that one
considers the situation where the kaon is emitted either in forward or 
in backward direction. Then we can look at two different cases:
\begin{itemize}
\item $\epsilon_d ||\hat p$. As real photons are transverse
  ($\lambda_\gamma = \pm 1$), $A^s$ and $C^{(ij)\, s}$ do not contribute. Thus
  for real photons and longitudinal target polarization ($\lambda_d = 0$),
  only the spin-triplet
  state contributes through $\vec b^t_1$ and  $C^{(ij)\, t}$.
  {\it Hence this is the case where we can study the spin-triplet final-state.}
  The observable that provides access to the longitudinal target polarization
  is
\be
\nonumber
(1-\sqrt{2}T^0_{20})\,\frac{d\sigma_0}{dm_{\Lambda n}d\Omega_q}
\sim |&b_1^t&+c_8^t+(c_6^t+c_7^t)(\hat{p}\vec{q}\,) \\
&+&c_5^t\vec{q}^{\,2}|^2,
\ee
where $T^0_{20}$ is defined by
\be
\nonumber
T^0_{20}
&=&\sum_{\lambda_1 , \lambda_2,
  \lambda_3}\Omega_{20;\lambda_1\lambda_3}M_{\lambda_1  \lambda_2}M^*_{\lambda_3  \lambda_2}
\\ 
&&/ \sum_{\lambda_1 , \lambda_2}M_{\lambda_1  \lambda_2}M^*_{\lambda_1  \lambda_2}
\ee
with $\lambda_1 , \lambda_3$ being the deuteron spin projection onto the photon 
momentum and $\lambda_2, \lambda_4$ the circular polarization
of the photon. The operators $\Omega_{ij}$ are defined by
\be
\nonumber
\Omega_{10}  = \sqrt{\frac32}
\begin{pmatrix}
1 & \phantom{-}0 & \phantom{-}0 \\
0 & \phantom{-}0 & \phantom{-}0 \\
0 & \phantom{-}0 & -1 \\
\end{pmatrix} , \
\Omega_{20} = \sqrt{\frac12}
\begin{pmatrix}
1 & \phantom{-}0 & \phantom{-}0 \\
0 & -2 & \phantom{-}0 \\
0 & \phantom{-}0 & \phantom{-}1 \\
\end{pmatrix} , \
\ee
\be
\nonumber
\Omega_{22} = \sqrt{3}
\begin{pmatrix}
0 & \phantom{-}0 & \phantom{-}1 \\
0 & \phantom{-}0 & \phantom{-}0 \\
0 & \phantom{-}0 & \phantom{-}0 \\
\end{pmatrix}, 
\Omega_{2-2} = \sqrt{3}
\begin{pmatrix}
0 & \phantom{-}0 & \phantom{-}0 \\
0 & \phantom{-}0 & \phantom{-}0 \\
1 & \phantom{-}0 & \phantom{-}0 \\
\end{pmatrix}, 
\ee
\be
\Omega^{c} \;  = \;
\begin{pmatrix}
1 & \phantom{-}0  \\
0 & -1
\end{pmatrix}, \;
\Omega^{l} \;  = \;
\begin{pmatrix}
\phantom{-}0 &-1  \\
-1 &\phantom{-} 0
\end{pmatrix} \ , 
\ee
see, e.g., Ref. \cite{Ohlsen1972}.
A complete description of the polarization observables for such kind
of reactions can be found in Ref. \cite{Arenhovel2005}.

\item $\vec \epsilon_d \perp \vec p$. In this case $C^{(ij)\, t}$
  vanishes, and the spin-singlet amplitudes (proportional to $A_s$ and $C^{(ij)\, s}$)
  are symmetric
  with respect to an interchange of $\vec \epsilon_d$ and $\vec \epsilon_\gamma$, whereas
  the spin-triplet amplitude (proportional to $\vec B^t$) is antisymmetric.
  This allows to construct combinations of spin observables containing only
  spin-singlet or spin-triplet contributions (two combinations for each spin), namely
\be
\nonumber 
(2+\sqrt{2} T_{20}^0&-&\sqrt{3} (T_{22}^l+T_{2-2}^l))\frac{d\sigma_0}{dm_{\Lambda n}d\Omega_q}
\\ \nonumber 
&=&-\sqrt{3}(\sqrt{2} T_{10}^c+ (T_{22}^l+T_{2-2}^l))\frac{d\sigma_0}{dm_{\Lambda n}d\Omega_q}\\
&\sim&\left|-a^s+\frac23 (c_1^s+c_3^s(\hat{p}\vec{q}\,)+c_2^s\vec{q}^{\,2})\right|^2 
\\ \nonumber 
(2+\sqrt{2} T_{20}^0&+&\sqrt{3} (T_{22}^l+T_{2-2}^l))\frac{d\sigma_0}{dm_{\Lambda n}d\Omega_q}
\\ \nonumber
&=&\sqrt{3}(-\sqrt{2} T_{10}^c+ (T_{22}^l+T_{2-2}^l))\frac{d\sigma_0}{dm_{\Lambda n}d\Omega_q}\\
&\sim& |b_1^t+b_2^t+b_3^t(\hat{p}\vec{q}\,)+b_4^t\vec{q}^{\,2}|^2,
\ee
with
\be
\nonumber T^l_{2\pm 2}
=&&\sum_{\lambda_1 , \lambda_2, \lambda_3,
 \lambda_4}\Omega_{2\pm 2;\lambda_1\lambda_3}\Omega^l_{\lambda_2\lambda_4}
M_{\lambda_1  \lambda_2}M^*_{\lambda_3  \lambda_4}
\\ \nonumber
&&/ \sum_{\lambda_1 , \lambda_2}M_{\lambda_1  \lambda_2}M^*_{\lambda_1  \lambda_2}\\
\nonumber
T^c_{10}
=&&\sum_{\lambda_1 , \lambda_2, \lambda_3,
  \lambda_4}\Omega_{10;\lambda_1\lambda_3}\Omega^c_{\lambda_2\lambda_4}
M_{\lambda_1  \lambda_2}M^*_{\lambda_3  \lambda_4}
\\ 
&&/ \sum_{\lambda_1 , \lambda_2}M_{\lambda_1  \lambda_2}M^*_{\lambda_1  \lambda_2}.
\ee

Here the
upper index ($c$ or $l$) refers to circularly or linearly polarized photons. Therefore,
the only possibility to obtain a pure spin-singlet $\Lambda N$ final-state is to 
perform a double polarization experiment.
\end{itemize}

%%%%%%%%%%%%%%%%%%%%%%%%%%%%%%%%%%%%%%%%%%%
\section{Influence of the meson-baryon interaction}
\label{influence}

\begin{figure}[h]
\begin{center}
\epsfig{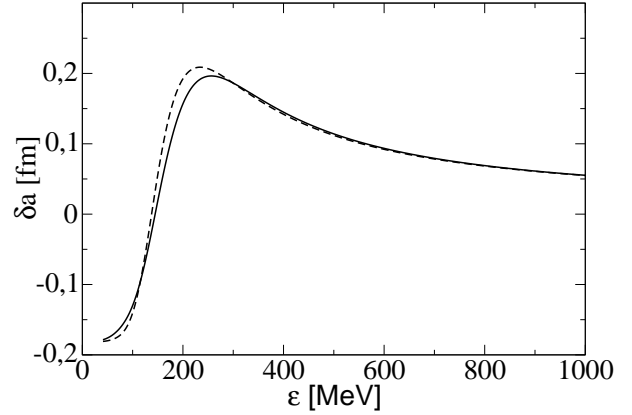}
\end{center}
\caption{Error in the extracted scattering length due to the presence of a resonance structure 
in the $K\Lambda$ subsystem depending on the available excess energy $\epsilon$. 
The solid line shows the result of the dispersion integral while the dashed line corresponds 
to an approximation, cf. discussion in Sec. \ref{influence}.
}
\label{fig3}
\end{figure}

As mentioned in the Introduction, in the derivation of Eq. (\ref{final})
we assumed that the interactions in the other two-body subsystems in the
final state are small. This concerns the $K \Lambda$ and the $KN$ systems. 
The reason was that for excess energies in the order of 100 to 200 MeV, 
the kinetic energy in those subsystems is large and does not vary strongly 
with the relative $\Lambda N$ momentum, when the latter system is considered
near its threshold for the extraction of the $\Lambda N$ scattering length, 
with the relative $\Lambda N$ momentum, when the latter system is considered
near its threshold for the extraction of the $\Lambda N$ scattering length, 
and therefore the energy dependence of the 
production amplitude should not change significantly.
It was noted, however, in \cite{Sibirtsev2005} that one should still 
be cautious because of possible effects due to the presence of $N^*$ 
resonances in the $K \Lambda$ system. Therefore, in the following we are going 
to derive some qualitative estimates as to how large
the effect of such resonances can be for the extracted $\Lambda N$ scattering
length. Note that this issue is relevant for both reactions $\gamma d \to K Y N$ and
$pp\to KYN$, although the details might differ. 
In particular the relative importance of 
contributions from resonances and of the background will depend on the
specific reaction mechanisms.
We assume here that the energy dependence of the production amplitude is
modified by a factor
\be
\Phi=\frac{1}{M_{\Lambda K}^2-M_R^2+i\Gamma_RM_R},
\ee
averaged over the $\Lambda N$ c.m. angle, i.e. we consider only the resonance 
contribution but neglect a background.
Here $M_R$ and $\Gamma_R$ are the Breit-Wigner mass and width of the resonance, respectively.
In general a nontrivial interference of the resonance amplitude with the background 
can produce a stronger mass dependence of the
production amplitude. On the other hand the sum of all partial waves
in the $\Lambda K$ system will have the opposite effect. Therefore, we
believe that the above approximation is reasonable in order to estimate the 
uncertainty in the extracted $\Lambda N$ scattering length induced 
by the presence of resonances in the $\Lambda K$ system. 
For simplicity we consider only an $S$-wave resonance.
Expanding $\Phi$ in terms of the $\Lambda N$ c.m. momentum $p$ one gets
\be
\nonumber \Phi&\approx & \frac{1}{\Delta M^2+i\Gamma_RM_R-Cp^2-2kpx} \\
\nonumber 
&\approx & \frac{1}{\Delta M^2+i\Gamma_RM_R} \\ 
\nonumber 
&\times &\left(1+\frac{Cp^2+2kpx}{\Delta M^2+i\Gamma_RM_R}
+ \frac{4k^2p^2x^2}{(\Delta M^2+i\Gamma_RM_R)^2}\right), \\
&&
\label{approx1}
\ee
where $C=2+\frac{m_\Lambda-m_n}{m_n}\left(\frac{s-m_K^2-(m_\Lambda+m_n)^2}{2m_\Lambda(m_\Lambda+m_n)}+1\right)$,
$k$ is the kaon momentum in the c.m. system of $\Lambda N$ (at $p=0$), $x$ is the cosine 
of the angle between the kaon and $\Lambda$ in the same system and
\be 
\nonumber
\Delta M^2&=&M_{\Lambda K}^2(p=0)-M_R^2 
=m_K^2+m_\Lambda^2 + \\
&& 
\frac{(s-m_K^2-(m_\Lambda+m_N)^2)m_\Lambda}{m_\Lambda+m_N}-M_R^2 \ .
\ee
Note that the factor $C$ is of the order of $2$ for the considered excess energies up to 
several hundred MeV because the mass difference between the $\Lambda$ and nucleon is 
small.
In order to estimate the effect that resonances may have on the extraction 
of the scattering length we evaluated the dispersion integral Eq.~(\ref{final})
for the amplitude given in Eq.~(\ref{approx1}). Corresponding results are shown in 
Fig.~\ref{fig3} as a function of the excess energy. Obviously, the resulting 
scattering length should be identical to
zero if there is completely no influence. We see that the deviations due to such
resonances are somewhat dependent on the excess energy but amount to $\pm 0.2$ fm 
at most. 

For a qualitative understanding of the role of the various scales it is
instructive to proceed as follows. 
Averaging over $x$ and removing a constant prefactor one obtains
\be
\nonumber 
<\Phi>&\sim& 1+\frac{Cp^2}{\Delta M^2+i\Gamma_RM_R}
+ \frac{4k^2p^2}{3(\Delta M^2+i\Gamma_RM_R)^2} \ .  \\ 
&&
\label{approx2}
\ee
For the production amplitude squared one gets the following 
mass dependence
\be
\nonumber
|A|^2\sim 1&+&\frac{2Cp^2\Delta M^2}{(\Delta M^2)^2+\Gamma_R^2M_R^2}
\\
&+& \frac{8k^2p^2((\Delta M^2)^2-\Gamma_R^2M_R^2)}{3((\Delta M^2)^2+\Gamma_R^2M_R^2)^2}.
\label{approx3}
\ee
The corresponding contribution to the $\Lambda N$ scattering length is 
(see Ref. \cite{Gasparyan2003}) 
\be
\nonumber
\delta a &\sim& p_{max}\left[\frac{C\Delta M^2}{(\Delta M^2)^2+\Gamma_R^2M_R^2}
+ \frac{4k^2((\Delta M^2)^2-\Gamma_R^2M_R^2)}{3((\Delta M^2)^2+\Gamma_R^2M_R^2)^2}\right],
\\
&&
\label{delta_a}
\label{approx4}
\ee
where $p_{max} \approx 200$ MeV/c reflects the limit of the dispersion integral.
It is easy to see that the result depends on two important scales: 
the resonance width (typically $150-200$ MeV)
and $\Delta M^2$ which is determined by the excess energy.
In order to obtain a rough idea for the order of magnitude of the corrections 
to the scattering length let us put $\Delta M^2=0$. Then
$\delta\, a \sim -4k^2p_{max}/(3\Gamma_R^2M_R^2)$.
If we take as a typical example the mass of the resonance to be $M_R=1700$ MeV
and its width to be $\Gamma_R=150$ MeV, then $k\sim 400$ MeV/c, which yields
$\delta\, a \sim -0.1$ fm. The absolute value of $\delta\, a$
becomes smaller as $\Delta M^2$ increases.
In Fig.~\ref{fig3} the result for $\delta\, a$ calculated 
by means of Eq.~\eqref{delta_a} is compared to the value obtained from 
the full dispersion integral Eq.~(\ref{final}). 
The two curves turn out to be almost identical and, therefore, justify 
the use of our approximations made in Eqs.~\eqref{approx1}-\eqref{approx4}. 
We conclude that for such excess energies where the available phase
space for the $K \Lambda$ system covers the resonance region one has
to expect an additional uncertainty of the extracted scattering length
in the order of $0.2$ fm---this has to be added to the one estimated
previously leading to a total theoretical uncertainty of $0.5$ fm. A
more concrete quantitative statement can be made only by analyzing the
actual experimental Dalitz plots, where one should clearly see whether
there is a strong dependence of the production amplitude on the $K
\Lambda$ invariant mass or not.

%%%%%%%%%%%%%%%%%%%%%%%%%%%%%%%%%%%%%
\section{Results and discussion}
\label{results}

As already said in the Introduction, the reaction $\gamma d \to K^+\Lambda n$
satisfies formally the main condition
needed for the dispersion integral method to be applied:
The momentum transfer in this reaction is large compared to 
the typical range of the final state $YN$ interaction. The required c.m.
momentum of the initial photon to produce the $YNK$ system at threshold
is equal to $584$ MeV/c. In order to be able to resolve the structure
induced by the $\Lambda N$ interaction one needs at least data covering
$\Lambda N$ invariant masses in the range of 40 MeV from the 
threshold, cf. Ref. \cite{Gasparyan2003}. In that work we argued also 
that the pertinent experiments should be performed preferably at somewhat 
higher total energies. Then there
will be no distortion of the signal within that 40 MeV range by the 
(upper) limit of the available phase space. Moreover, effects from possible 
interactions in the other final states ($KN$, $K\Lambda$) are 
kinematically better separated and should not influence the results
for $\Lambda N$ too much. 

An important kinematical constraint for the reaction $\gamma d \to K^+\Lambda n$
is the limitation of the kaon angle to very forward or very backward 
directions because only then a separation of the spin-singlet and spin-triplet 
states is possible, as shown in Sec. \ref{observables}. 
Unfortunately, there are indications that the total count rate could
be very small in the backward region. For example, 
the model calculation of Salam and Arenh\"ovel~\cite{Salam} suggests
that the cross sections drop dramatically in that angular range, cf. 
their Fig. 12. 
This can be easily understood within the impulse approximation. In this
case the spectator nucleon carries necessarily a large momentum for 
kaons produced in backward direction and for such large momenta the
deuteron wave function is strongly suppressed. Additional 
production mechanisms that involve two-step processes, considered also
in Ref.~\cite{Salam}, relax the situation somewhat. But still  
it could be difficult to perform measurements for the backward region 
and one has to wait for concrete experiments in order to see whether 
sufficient statistics can be achieved. 

\begin{figure}[h]
\begin{center}
\epsfig{file=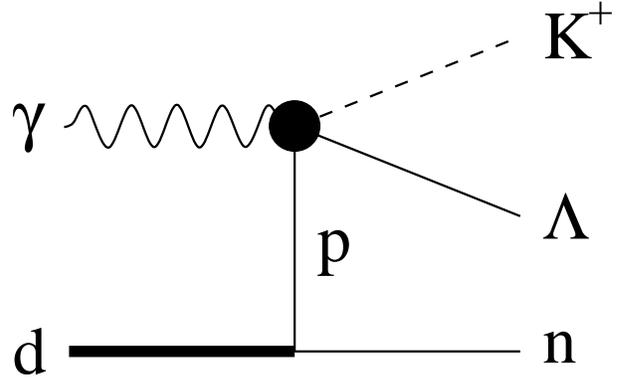, width=8.0cm}
\end{center}
\caption{Diagram corresponding to the quasi-free kaon production on the proton.}
\label{fig1}
\end{figure}
 
Therefore, in the following we will concentrate on results for forward
angles. However, in this case there is a particular singularity
of the production amplitude that imposes some restrictions on the application 
of our method. It is the so-called
quasi-free production mechanism (see Fig. \ref{fig1}). When the available excess
energy in the $\Lambda n K^+$ system is around $90$ MeV or more then the production of the 
$\Lambda K^+$ system is possible on a single proton, resting in the deuteron rest frame.
Therefore, this effect introduces a new, very 
small scale in the production operator caused by the small deuteron binding energy.
It is clear that this particular production mechanism is dominant when the 
relative momentum of the two nucleons inside the deuteron is not large. 
Thus, it influences primarily forward kaons where then the $\Lambda N$ system is 
moving in direction of the deuteron momentum in the c.m. system. 
Note that the peak is shifted
somewhat away from very low relative $YN$ momenta because the photon cannot produce 
a $\Lambda$ at rest on a proton at rest. 

\begin{figure}[h]
\begin{center}
\epsfig{file=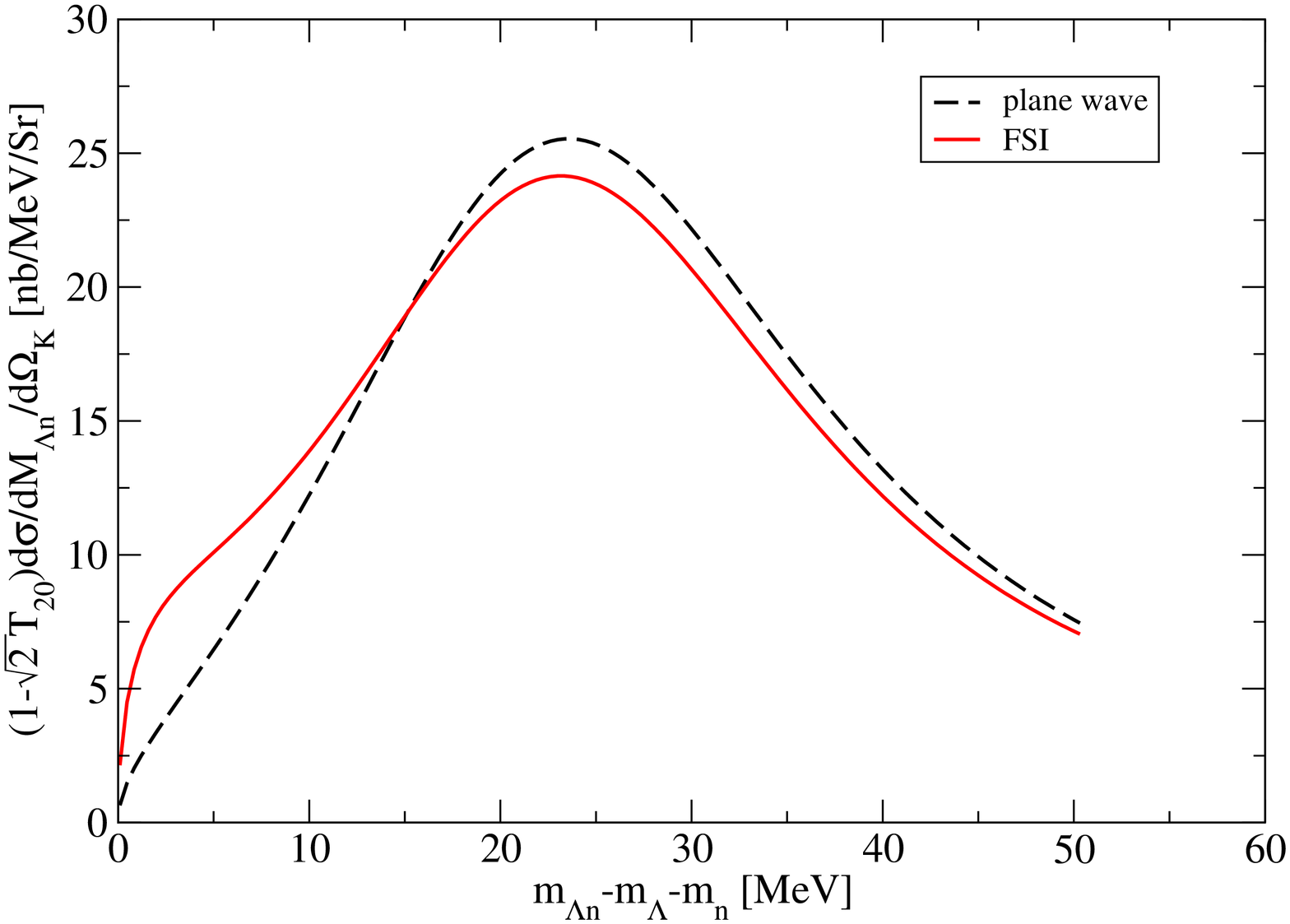, width=6.5cm, angle=-90} 
\epsfig{file=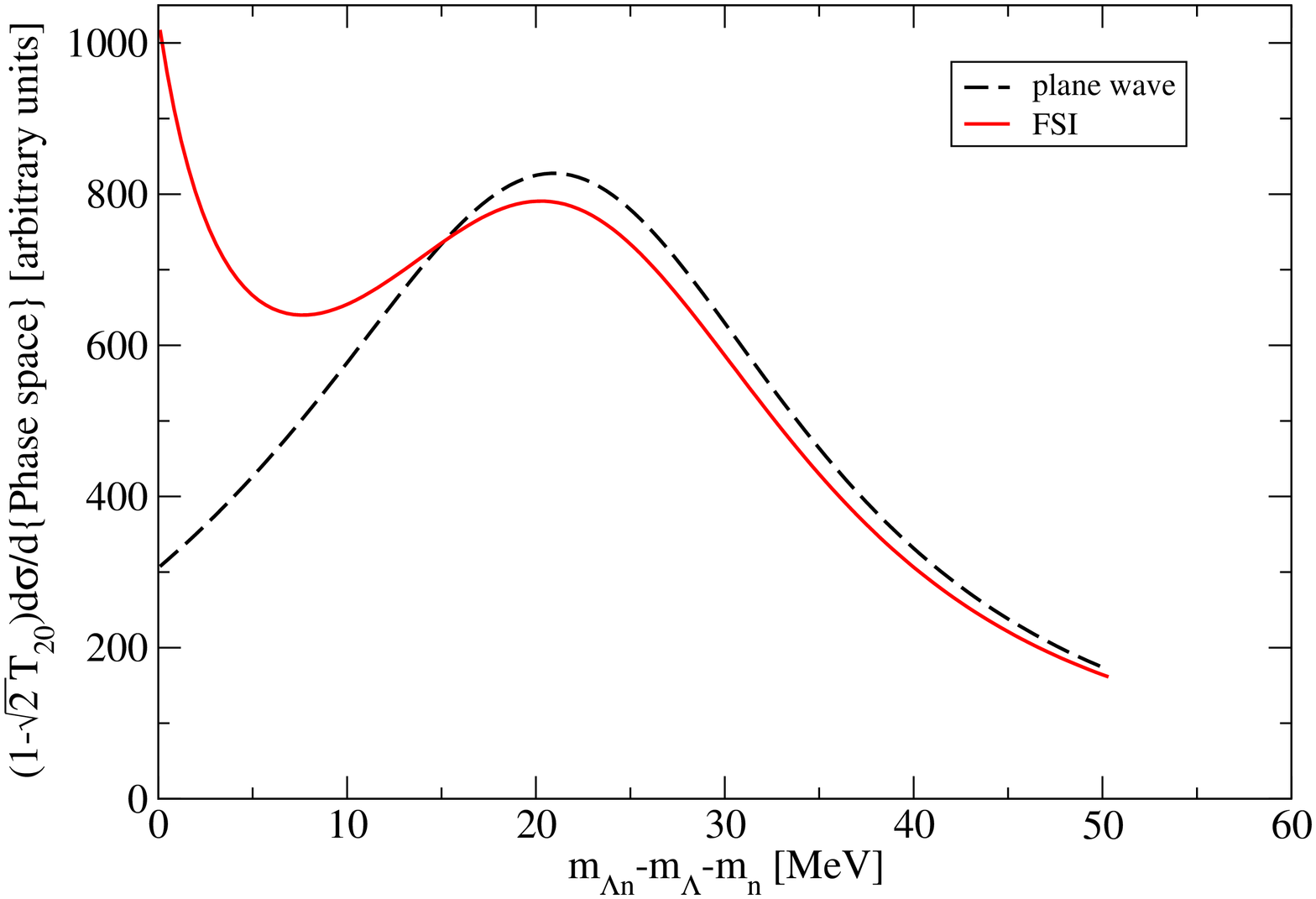, width=6.5cm, angle=-90} 
\end{center}
\caption{Top: Model results for the spin-dependent observable $1-\sqrt{2} T_{02}^0$ at 
$E_\gamma$ = 1300 MeV and $\Theta_K = 0^o$ as a function of the 
$\Lambda n$ invariant mass $m_{\Lambda n}$. The dashed line is the
impulse approximation while the solid line is the full result including
the $\Lambda n$ final-state interaction. 
Bottom: Same results but the phase-space factor is diveded out. 
}
\label{figT20_1300}
\end{figure}

In Fig.~\ref{figT20_1300} we demonstrate the situation for a concrete model 
calculation where results for $1-\sqrt{2}T_{02}^0$ at $E_\gamma$ 
= 1300 MeV ($\epsilon$ = 349 MeV) are shown for kaon production in forward direction 
($\Theta_K= 0^o$). 
Details of the model calculation can be found in Ref.~\cite{Yama}. Let us
mention here that the calculation is done in the impulse approximation including
the $YN$ final-state interaction, utilizing 
the deuteron wave function of the Nijmegen93 potential \cite{Nij93} and
the NSC97f $YN$ force \cite{Stoks}. The elementary kaon-production amplitude on the 
nucleon ($\gamma N\to K\Lambda$) is derived from a set of tree-level Feynman diagrams 
where the free parameters have been fixed 
so that all available $K^+\Lambda$, $K^+\Sigma^0$, and $K^0\Sigma^+$ 
photoproduction data in the relevant energy region are reproduced \cite{Bennhold}. 
Additional production mechanisms involving, e.g., $KN$ rescattering or the
$\pi N \to K\Lambda$ process, considered in Ref.~\cite{Salam}, are not included
in this model. However, those mechanisms contribute predominantly for kaon production
at backward angles \cite{Salam} and are not so important for the forward angles
we consider. 

The model calculation presented in Fig.~\ref{figT20_1300} clearly shows the 
presence of a bump due to quasi-free kaon production. It occurs at fairly
small $\Lambda n$ invariant masses and, therefore, makes a reliable determination
of the $\Lambda n$ scattering length from data impossible. 
Thus, for extracting the $\Lambda n$ scattering length from forward-angle data
one has to consider the reaction $\gamma d \to K^+\Lambda n$ for energies below the
appearance of this quasi-free peak, i.e. at excess energies $40-50$ MeV. 
First of all one should note that 
the influence of the $\Lambda K$ interaction is not necessarily much 
stronger then at higher excess energies, since we are within the resonance
region in both cases. Therefore, the uncertainty of the method remains the same.
This issue was addressed already in the previous section.
Another problem is the limited phase space at low excess energies.
The phase space is proportional to $q' \times p' \times dm_{\Lambda n}$. 
Since we are interested in the region of small relative momenta $p'$ 
in the $\Lambda n$ system in any case the suppression enters only 
due to the factor $q'$. The concrete effect of the suppression 
depends, of course, on the actual shape of the mass spectrum, but to get 
a rough estimate one can compare the $q'$ values for different
excess energies at the $\Lambda n$ threshold ($p'=0$). 
For example, for the excess energy $50$ MeV this value is about $2.5$ 
times smaller than for $300$ MeV. This means that the 
suppression is not such a serious problem in our case.
 
An interesting side aspect at low excess energies is that then 
also the kaons should be predominantly produced in an $S$-wave relative to 
the $\Lambda n$ system (unless the quasifree mechanism remains
dominant even at low energies).  
In a pure $S$ wave situation the allowed
operator structure given in Eqs.~(\ref{singlet},\ref{triplet}) 
simplifies significantly and, in particular, the reaction amplitude does 
not depend on the direction of the kaon momentum anymore. Consequently, 
all expressions in Sec.~\ref{observables} are valid for arbitrary 
angles. Therefore, one can work with observables integrated over 
the kaon angle in the c.m. system which means that a significant 
enhancement of the experimental statistics can be achieved. 
In addition the angular integration allows to get rid of interference terms 
between the $S$- and (small) $P$-waves that depend linearly on the
kaon momentum so that possible influences from the energy dependence of 
the production operator, which is primarily due to terms linear in $q'$,
are minimized. 

\begin{figure}[h]
\begin{center}
\epsfig{file=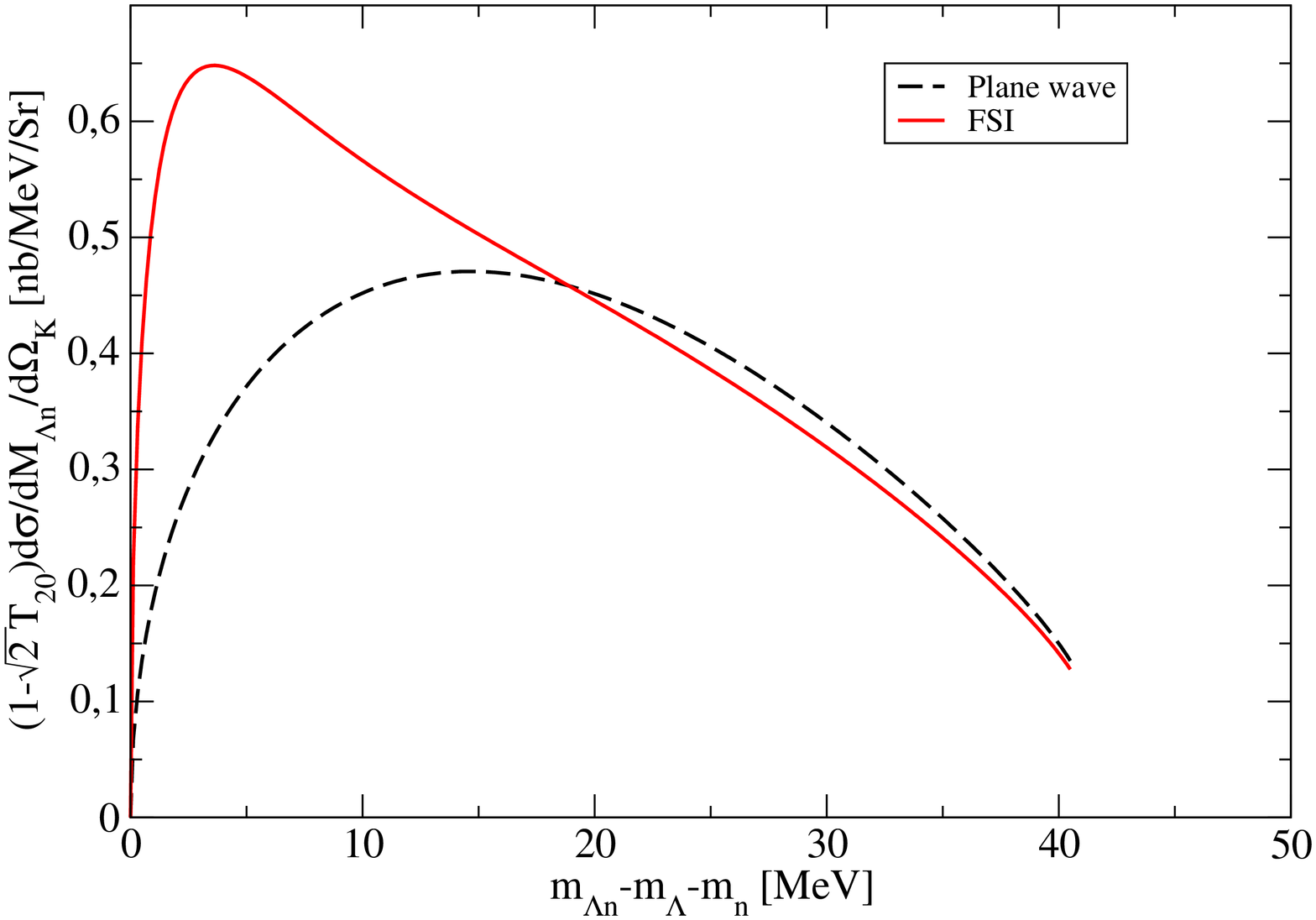, width=6.5cm, angle=-90} 
\epsfig{file=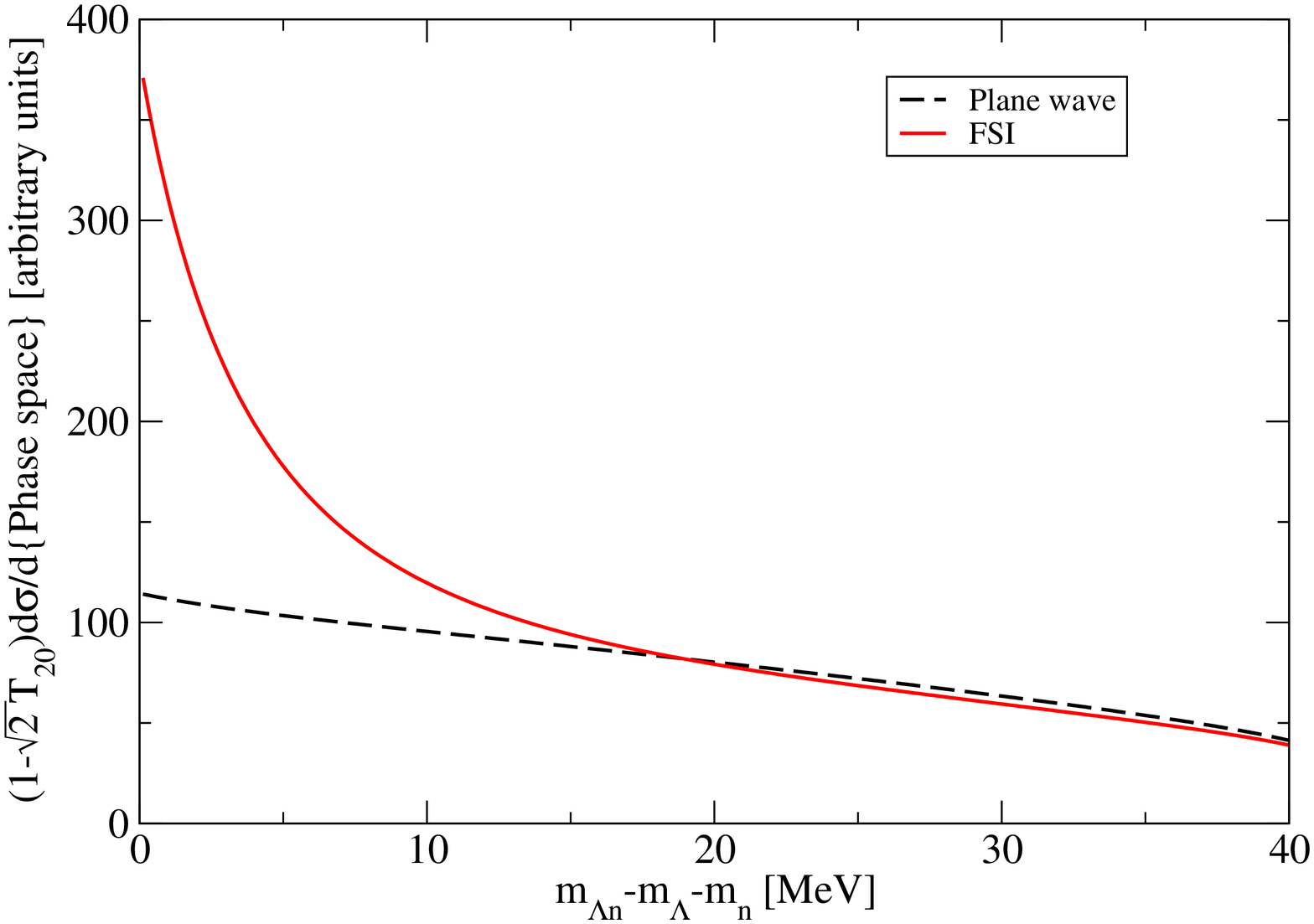, width=6.5cm, angle=-90} 
\end{center}
\caption{Top: Model results for the spin-dependent observable $1-\sqrt{2} T_{02}^0$ at 
$E_\gamma$ = 850 MeV and $\Theta_K = 0^o$ as a function of the 
$\Lambda n$ invariant mass $m_{\Lambda n}$. The dashed line is the
impulse approximation while the solid line is the full result including
the $\Lambda n$ final-state interaction. 
Bottom: Same results but the phase-space factor is diveded out. 
}
\label{figT20}
\end{figure}

In Fig.~\ref{figT20} we show predictions of the model calculation \cite{Yama}
for the spin observable $1-\sqrt{2}T_{02}^0$ for $E_\gamma$ = 850 MeV 
($\epsilon$ = 41.5 MeV) and forward kaons. The dashed line is the 
result for the impulse approximation while the solid line corresponds to
the full model including the $\Lambda N$ FSI. It is obvious how strongly
the $\Lambda N$ interaction modifies the observable for invariant masses 
close to the $\Lambda N$ threshold. When applying the dispersion integral
method to this observable, cf. Ref.~\cite{Gasparyan2003} for details, 
we obtain the scattering length of -2.06 $fm$ for the $^3S_1$ partial wave.
This has to be compared with $a_t$ = -1.70 $fm$ of the $YN$ model \cite{Stoks}
used for the model calculation. Thus, the extracted scattering
length differs from the one utilized in the model calculation by about
0.4 $fm$, which is in line with the uncertainty that is expected for 
the method \cite{Gasparyan2003}. Specifically, one has to keep in mind that the 
present model calculation includes also the uncertainties discussed in 
Sec. \ref{influence} because it is based on an elementary kaon-production 
amplitude that involves resonances in the $\Lambda K$ channel \cite{Bennhold}. 

\section{Summary}
\label{summary}

In the present paper we have studied the perspects of utilizing the 
strangeness-production reaction $\gamma d \to K^+\Lambda n$
for the determination of the $\Lambda n$ low-energy scattering parameters. 
In particular, we derived those 
spin observables that need to be measured in order to isolate the $\Lambda n$ 
singlet ($^1S_0$) and triplet ($^3S_1$) states, 
and we presented concrete results for one of those observables based on
a model calculation by Yamamura et al.~\cite{Yama}. 

It turned out that a separation of the singlet and triplet states is 
feasible for experiments with kaons emitted either in forward or backward 
direction. On the other hand we found that the quasi-free production process, 
which dominates the reaction in forward direction at higher energies
($E_\gamma \geq 900$ MeV/c), distorts
the $\Lambda N$ invariant mass spectum so strongly that this particular
kinematics cannot be used to extract the $\Lambda N$ scattering lengths
reliably. However, the situation looks very promising for experiments
for energies just below the appearance of this 
quasi-free peak ($E_\gamma \approx 850$ MeV/c). For this kinematics we 
presented a test calculation 
where we generated the required spin observables from the model of 
Yamamura et al.~\cite{Yama} and then we applied to them the 
dispersion integral method for extracting the $\Lambda n$ 
scattering length. The value obtained for the $^3S_1$ scattering
length differs from the one utilized in the model calculation by about
0.4 $fm$, i.e. lies within the uncertainty that is expected for the
method \cite{Gasparyan2003}. A determination of the scattering
lengths is also possible from data at backward angles at any energy. 
However, for backward kaon production all model calculations predict rather 
small count rates. Thus, one was to wait for concrete experiments in order
to see whether sufficient statistics can be achieved for this kinematics. 

The presented estimation of the uncertainties of the extracted $\Lambda N$
scattering length that could arise from the interaction in the other
final states ($K\Lambda$, $KN$), together with the results of 
a concrete application, implies that the reaction $\gamma d \to K^+\Lambda n$ 
could allow to determine the $\Lambda N$ scattering lengths with an accuracy 
similar to the reaction $pp\to K^+\Lambda p$.
Thus, we believe that the photon-induced reaction is an interesting 
alternative for extracting the $\Lambda N$ scattering lengths and it is also 
very useful for cross-checking results obtained from the purely hadronic 
strangeness production. 

\subsection*{Acknowledgments}
A.G. thanks the Institut f\"ur Kernphysik at the
Forschungs\-zentrum J\"ulich for its hospitality during the period when
the present work was carried out.
Furthermore, he would like to acknowledge financial support by the
grant No. 436 RUS 17/75/04 of the Deutsche Forschungsgemeinschaft
and by the Russian Fund for Basic Research, grant No.06-02-04013.

\end{document}